\documentclass[12pt]{article}
\catcode`\@=11 \@addtoreset{equation}{section}

\usepackage{amsbsy,amssymb,latexsym,amsfonts,amsmath,cite}
\usepackage{color}

\allowdisplaybreaks

\global\arraycolsep=2 pt
\newcommand\diag{\mathrm{diag}}
\newcommand\CN{\mathcal N}
\def\tr{\mathop{\rm tr}}

\newcommand\CL{{\mathcal L}}

\newcommand{\adss}[2]{{AdS$_{#1}\times$S$^{#2}$}}
\newcommand{\ads}[1]{{AdS$_{#1}$}}
\newcommand{\s}[1]{{S$^{#1}$}}
\newcommand\cp[1]{$\mathbb{C}$P$^{#1}$}
\newcommand{\adssz}[3]{{AdS$_{#1}\times$S$^{#2}/\bbZ_{#3}$}}

\newcommand\BL{\textbf{L}}

\newcommand\bbZ{{\mathbb Z}}

\newcommand\adscp{AdS$_4\times\mathbb{C}$P$^{3}${}}

\begin{document}

\begin{flushright}
OIQP-10-02 \\
CQUeST-2010-0379 \\
KUNS-2281 \\
{RIKEN-TH-188}
\end{flushright}

\vspace*{0.5cm}

\begin{center}
{\Large \bf 
Semiclassical Analysis of M2-brane in \adssz{4}{7}{k}
}
\end{center}
\vspace{10mm}

\centerline{\large 
Makoto Sakaguchi$^1$~, 
Hyeonjoon Shin$^{2,3}$~, and
Kentaroh Yoshida$^4$
}

\vspace{8mm}

\begin{center}
$^1${\it Theoretical Physics Laboratory, RIKEN\\ Wako, Saitama 351-0198, Japan}\\
{\tt msakaguchi@riken.jp}
\vspace{5mm}

$^2${\it Center for Quantum Spacetime\\
Sogang University, Seoul 121-742, South Korea}\\

$^3${\it Department of Physics, 
 Pohang University of Science and Technology,\\
 Pohang 790-784, South Korea}\\
{\tt hshin@sogang.ac.kr, hyeonjoon@postech.ac.kr}
\vspace{5mm}

$^4${\it Department of Physics, Kyoto University\\
Kyoto 606-8502, Japan}\\
{\tt kyoshida@gauge.scphys.kyoto-u.ac.jp}

\end{center}

\vspace{1cm}

\begin{abstract}
  We start from the classical action describing a single M2-brane on
  AdS$_4\times$S$^7/\mathbb{Z}_k$ and consider semiclassical
  fluctuaitions around a static, 1/2 BPS configuration whose shape is
  AdS$_2\times$S$^1$. The internal manifold S$^7/\mathbb{Z}_k$ is
  described as a U(1) fibration over $\mathbb{C}$P$^3$\ and the static
  configuration is wrapped on the U(1) fiber. Then the configuration 
  is reduced to an AdS$_2$ world-sheet of type IIA string on
  AdS$_4\times\mathbb{C}$P$^3$ through the Kaluza-Klein reduction on 
  S$^1$.  It is shown that the fluctuations form an infinite set of 
  $\mathcal{N}=1$ supermultiplets on AdS$_2$\,  {for $k=1,2$}. 
  The set is invariant under SO(8) which may be consistent with 
  $\mathcal{N}=8$ supersymmetry on AdS$_2$.
  We discuss the behavior of the fluctuations around the boundary of
  AdS$_2$ and its relation to deformations of Wilson loop operator.
\end{abstract}

\thispagestyle{empty}
\setcounter{page}{0}
\newpage

\section{Introduction}

More than a decade has passed from the discovery of the AdS/CFT
correspondence \cite{AdS/CFT,GKPW}.  It is still giving important
arenas to study many aspects of string theory.  
Depending on the dimensionality, one can have various versions
of AdS/CFT.  Undoubtedly, AdS$_5$/CFT$_4$ may be the most widely studied case.  As for other
cases, however, especially for AdS$_4$/CFT$_3$\,,
there had been not so much progress because it seemed very difficult
to construct a three-dimensional interacting $\mathcal{N}=8$
superconformal field theory which describes the low energy dynamics on
multiple M2-branes \cite{Schwarz}.

\medskip 

A couple of years ago, one possible resolution for the problem of constructing such a field theory
has been provided by Bagger, Lambert \cite{BL} and Gustavsson \cite{G}
(BLG) by utilizing the Lie 3-algebra. Although their works known as
the BLG theory contain the unusual 3-algebra structure, the theory is
in fact equivalent to the conventional gauge theory as shown in
\cite{VanRaamsdonk:2008ft}. The BLG theory has triggered off a
break-through in the study of AdS$_4$/CFT$_3$\,, and the equivalence
of it with the conventional gauge theory has led Aharony, Bergman,
Jafferis and Maldacena (ABJM) \cite{ABJM} to propose the duality
between type IIA string theory on AdS$_4\times\mathbb{C}$P$^3$ and
$\mathcal{N}=6$ superconformal Chern-Simons matter system in three
dimensions. A numerous amount of the works related to this issue have been carried
out so far, and hence our understanding on AdS$_4$/CFT$_3$ 
is now making considerable progress. 

\medskip 

By the way, an important check of the AdS/CFT duality is the correspondence 
between a 1/2 BPS Wilson loop and a string world-sheet whose shape is AdS$_2$\cite{Wilson1,Wilson2}. 
It is well studied in the case of AdS$_5$/CFT$_4$ and it is an important observation 
that the exact expectation value of the 1/2 BPS circular Wilson loop can be computed  
by a Gaussian matrix model \cite{GM1,GM2,GM3}. 
(For other works on supersymmetric Wilson loops, for example, see \cite{SW1,SW2}.)

\medskip 

On the other hand, it is not so obvious to understand Wilson loops in AdS$_4$/CFT$_3$\,. 
There are 1/2 BPS AdS$_2$ solutions in type IIA string theory 
on AdS$_4\times\mathbb{C}$P$^3$\,. On the other hand,
in the $\mathcal{N}=6$ Chern-Simons matter system, it is possible to
construct at most 1/6 BPS Wilson loops \cite{DPY,Chen,RSY} as far as
the bosonic degrees of freedom are taken into account.  It has been recently shown
that 1/2 BPS Wilson loops can be constructed by including the
fermionic degrees of freedom \cite{supermatrix}
(see \cite{Lee} also). However, it is not
clear why the fermionic contributions are necessary to realize the 1/2
BPS configuration while the corresponding 1/2 BPS string solution is
purely bosonic. Another related question is what is the 
eleven-dimensional origin of the fermionic contributions to the 1/2 
BPS Wilson loops.

\medskip 

Motivated by these questions, in this paper we will consider a
semiclassical approximation of a single M2-brane on
AdS$_4\times$S$^7/\mathbb{Z}_k$ background\footnote{For semiclassical 
strings on AdS$_5\times$S$^5$\,, see for example the excellent reviews 
\cite{revs} and the references therein. Semiclassical approximation
for Wilson loops was originally discussed 
in \cite{DGT} and the quadratic action obtained there coincides
with the non-relativistic action \cite{GGK,SY}. 
The correspondence between the semiclassical approximation
and the non-relativistic limit 
was confirmed also for the AdS-brane cases \cite{n3}.  }. 
For this purpose we begin with the classical
action describing a single M2-brane on AdS$_4\times$S$^7/\mathbb{Z}_k$
and expand it around a static, 1/2 BPS classical solution.  This
solution\footnote{The solution was 
originally argued in \cite{Wilson1}. It is also discussed in classifying 1/2 BPS AdS-branes 
\cite{Kim-Yee,SY:NH,Lunin:2007ab}.}  
has the shape of AdS$_2\times$S$^1$\,. The internal manifold
S$^7/\mathbb{Z}_k$ may be described as a U(1) fibration over
$\mathbb{C}$P$^3$ and the S$^1$ part of the solution is wrapped on the 
U(1) fiber. Then the M2-brane solution is reduced to an AdS$_2$ 
world-sheet of type IIA string on AdS$_4\times\mathbb{C}$P$^3$ through 
the  Kaluza-Klein (KK) reduction on the S$^1$\,. We show that the fluctuations form 
an infinite set of $\mathcal{N}=1$ supermultiplets on AdS$_2$\, for $k=1,2$
and that the set is
composed of $\mathcal{N}=8$ supermultiplets on AdS$_2$.  Finally we
discuss what kind of the fluctuations can reach the boundary consistently.  


\medskip 

This paper is organized as follows. In section 2, we introduce the
classical action of a single M2-brane on
AdS$_4\times$S$^7/\mathbb{Z}_k$\,. Our notation and convention are
also summarized. In section 3, we introduce a static, classical
M2-brane solution whose shape is AdS$_2\times$S$^1$ and discuss a 
semiclassical approximation around the solution.\footnote{
The resulting action has the same form as the non-relativistic M2-brane action derived in \cite{SY:NH}.
}  
In section 4, we
consider the KK reduction of the classical solution and the
semiclassical fluctuations around it. The resulting spectrum consists
of an infinite number of $\mathcal{N}=1$ supermultiplet on
AdS$_2$\, {for $k=1,2$}.
It is shown that the
$\mathcal{N}=1$ supermultiplets can be combined to form an infinite 
set of the $\mathcal{N}=8$ supermultiplet on AdS$_2$\,.  In section 5 we investigate 
the boundary behaviour of the fluctuations and identify the modes that can reach the boundary. 
The final section is devoted to conclusion and discussion.

\section{M2-brane action on \adss{4}{7}$/\bbZ_k$}

The starting point of our discussion is the classical action
describing a single M2-brane on AdS$_4\times$S$^7/\mathbb{Z}_k$\,. The
purpose of this section is to introduce it and summarize the notation
and convention utilized in this paper.

\subsection{AdS$_4\times$S$^7/\mathbb{Z}_k$ background}

%

The \adss{4}{7}$/\bbZ_k$ background is described by 
\begin{eqnarray}
  ds_{11}^2 &=& \frac{R^2}{4}ds^2_{\rm AdS_4} + R^2 ds^2_{{\rm S}^7/\mathbb{Z}_k}\,, \nonumber \\ 
  ds^2_{\rm AdS_4} &=& \frac{1}{z^2}(-dt^2+dx_1{}^2+dx_2{}^2+dz^2)\,,  
   \nonumber \\ 
  ds^2_{{\rm S}^7/\mathbb{Z}_k} &=& ds^2_{\mathbb{C}{\rm P}^3} + \frac{1}{k^2} (dy+kA)^2\,,
\label{bg}  
\end{eqnarray}
with which the four-form field strength is equipped: 
\begin{eqnarray}
F_4 = -\frac{3}{8}\frac{R^{3}}{z^4}\, 
dt\wedge dx_1\wedge dx_2 \wedge dz\,. 
\label{flux}
\end{eqnarray}
The radius of \ads{4} is given by $R_\mathrm{AdS}=R/2$ while that of
\s{7} is $R$\,. For notational convenience, we will set $R=1$ below. 
Note that 
the period of $y$ is $2\pi$\,, i.e., $y\sim y +2\pi$\,.

\medskip 

The space ${\rm S}^7/\mathbb{Z}_k$ in (\ref{bg}) is described
as a $U(1)$ fibration over $\mathbb{C}$P$^3$\,. 
The field $A$ is the one-form potential and leads to the 
K\"ahler form $F\equiv \frac{1}{2}dA$ on $\mathbb{C}$P$^3$\,. 
For the explicit expression of the K\"ahler form $F$, we adopt the
following one:
\begin{eqnarray}
F=\frac{i}{2}\partial\bar\partial \ln (1+w^m\bar w^m)
=\frac{i}{2}\frac{dw^m\wedge d\bar w^n}{(1+|w|^2)^2}\Big(
\delta_{mn}(1+|w|^2)
-\bar w^m w^n
\Big) \,,
\end{eqnarray}
where $w^m$ ($m=1,2,3$) are the complex coordinates on \cp{3}\,.  
Then the metric on $\mathbb{C}$P$^3$ is given by the 
Fubini-Study metric,
\begin{eqnarray}
ds^2_{\mathbb{C}{\rm P}^3} = \frac{(1+|w|^2)dw^nd\bar{w}^n 
- (\bar{w}^nw^mdw^nd\bar{w}^m)}{(1+|w|^2)^2}\,.
\end{eqnarray}



\medskip 

From the metric (\ref{bg}), the elfbein $e^A~(A=0,\ldots,9,\natural)$ is taken as
\begin{eqnarray}
e^a&=&\left( \frac{dt}{2z},\frac{dx_1}{2z},\frac{dx_2}{2z},\frac{dz}{2z}\right)~,~~~
e^{a'}=\left(e^{\tilde a},e^\natural=\frac{1}{k}(dy+kA)\right)~,
\label{elfb}
\end{eqnarray}
where the index $A$ is decomposed as $A=(a,a')=(a,\tilde{a},\natural)$
with the following ranges:
\begin{itemize}
\item \qquad $a=0,\ldots,3$ is the index for AdS$_4$\,, 
\item \qquad $\tilde{a}=4,\ldots,9$ is for $\mathbb{C}$P$^3$\,,
\item \qquad $a'=(\tilde{a},\natural)$ is for S$^7/\mathbb{Z}_k$\,, where 
$\natural$ denotes the $U(1)$ fiber direction. 
\end{itemize}
With the elfbein, the non-vanishing components of spin
connection $w^{AB}$ are evaluated as  
\begin{eqnarray}
&&  w^{03}=-\frac{dt}{z}\,, \qquad 
w^{13}=-\frac{dx_1}{z}\,, \qquad 
w^{23}=-\frac{dx_2}{z}\,,  \nonumber \\ 
&& w^{\tilde a\tilde b}=\hat w^{\tilde a\tilde b}-F^{\tilde a\tilde b} 
  e^\natural\,, \qquad 
w^{\natural \tilde a}=F^{\tilde a}{}_{\tilde b}e^{\tilde b}\,, 
\end{eqnarray}
and $\hat w^{\tilde a \tilde b}$ is the spin connection on \cp{3}.
Note here that $F^{\tilde a\tilde b}$ can be taken to 
be of the form
\begin{eqnarray}
F^{\tilde a\tilde b}=\left(
  \begin{array}{ccc}
  \varepsilon && \\ 
  &\varepsilon & \\
  &&\varepsilon 
  \end{array}
\right)\,, \qquad 
\varepsilon=\left(
  \begin{array}{cc}
  0 & 1 \\
  -1&0 
  \end{array}
\right)
\end{eqnarray}
without loss of generality.

\subsection{M2-brane action on AdS$_4\times$S$^7/\mathbb{Z}_k$}

Let us consider the classical action describing a single 
M2-brane on AdS$_4\times$S$^7/\mathbb{Z}_k$ (rather than multiple 
M2-branes)\,.\footnote{A similar analysis for the finite values of $k$ is discussed 
in \cite{Ahn:2008xm}, where spinning string solutions are considered in comparison to 
our static configuration.}  
Apart from the 
$\mathbb{Z}_k$ quotient, the corresponding action has already been 
constructed in \cite{m2ads} and one may consult it in this
subsection. 

\medskip 

The classical action of a single M2-brane is composed of the
Nambu-Goto (NG) action and the Wess-Zumino (WZ) term:
\begin{equation}
S_{\rm M2} = S_{\rm NG} + S_{\rm WZ}\,.
\end{equation}
Firstly, the NG action is given by 
\begin{eqnarray}
S_\mathrm{NG} &=& T\int\! d^3\xi\, \sqrt{-\det g}\,, \qquad 
g_{ij}=\BL_{i}^A \BL_{j}^B \eta_{AB}\,,  
\end{eqnarray}
where $T$ is the M2-brane tension and $\xi^i=(\tau,\sigma,\rho)$ are
the world-volume coordinates. The induced metric $g_{ij}$ is given 
in terms of the super elfbein
\begin{eqnarray}
\BL_{i}^A=\partial_i Z^{\hat M}\BL_{\hat M}^A\,,\qquad 
Z^{\hat M}=(X^M,\theta) \qquad (M=0,\ldots,9,\natural)\,.
\end{eqnarray}
In our notation of the eleven-dimensional superspace coordinate,
$Z^{\hat M}$,  $M$ labels the curved space-time 
coordinates, and $\theta$ is 
the 32-component Majorana spinor.  
The indices $A,~B,\ldots$ labels the local Lorentz frame.

\medskip

For our purpose, it is sufficient to consider the
super elfbein $\BL^A$ 
expanded explicitly  up to  
the second order of $\theta$:
\begin{eqnarray}
\BL^A &=& e^A-\bar\theta\Gamma^A D\theta+O(\theta^4)\,.  
\end{eqnarray}
Here $\Gamma^A$'s are the SO(1,10) gamma matrices 
and $\bar\theta=\theta^TC$ with the charge conjugation matrix $C$.
The covariant derivative for $\theta$ is defined as 
\begin{eqnarray}
D\theta &\equiv& d\theta
-e^aI\Gamma_a\theta
-\frac{1}{2}e^{a'}I\Gamma_{a'}\theta
+\frac{1}{4}w^{AB}\Gamma_{AB}\theta\,, \qquad I \equiv \Gamma^{0123}\,, 
\end{eqnarray}
where the second and third terms come from the coupling to the
four-form field strength $F_4$\,.

\medskip 

Another part of the M2-brane action, the WZ term, is given by
\begin{eqnarray}
S_\mathrm{WZ} &=& T\int\left[
{}^*c_{(3)}+\int_0^1\!dt~^*(-\hat\BL^A \wedge \hat\BL^B \wedge
  \hat{\bar L} \Gamma_{AB}\theta)
\right]\,, \\ 
dc_{(3)} &=& -\frac{6}{4!}\epsilon_{a_1\cdots a_4}
  e^{a_1} \wedge \cdots \wedge e^{a_4}\,, 
\label{dc3}
\end{eqnarray}
where $\hat \BL=\BL|_{\theta\to t\theta}$\,, $\hat L =L|_{\theta\to t
  \theta}$ and we assume that $L$ is defined up to the second order of $\theta$: 
\begin{eqnarray}
L &=& D\theta +O(\theta^3)\,. \nonumber 
\end{eqnarray} 
The $\ast$-operation means the pullback of the
background geometry to the world-volume of M2-brane.

\medskip

Let us now expand the M2-brane action up to the quadratic order in $\theta$\,. Since the background
of our concern does not have non-vanishing fermionic field, there are
no terms linear in $\theta$, and thus the expansion results in
\begin{eqnarray}
S_{\rm M2} &=&S_{B}+S_{F}+O(\theta^4) \,,
\end{eqnarray}
where $S_B$ is the purely bosonic part, that is, the zeroth order part in $\theta$,
and $S_F$ is the part of quadratic order.  Before presenting each part of the action, 
it is convenient to expand the metric in powers of $\theta$ as\footnote{
We use the following notation: 
$a_{(i} b_{j)}=\frac{1}{2}(a_{i} b_{j}+a_{j} b_{i})$
and
$a_{[i} b_{j]}=\frac{1}{2}(a_{i} b_{j}-a_{j} b_{i})$.
}
\begin{eqnarray}
g_{ij} &=& g_{B\,ij}+g_{F\,ij} +O(\theta^4)\,, \\ 
g_{B\,ij} &=& e^A_ie^B_j \eta_{AB}\,, \\
g_{F\,ij} &=& -2e^A_{(i}\bar\theta \Gamma^B D_{j)}\theta\eta_{AB}\,,
\end{eqnarray}
where $e^A_i \equiv \partial_i X^M e^A_M$.
Then the purely bosonic part is written down with $g_{B\, ij}$ and $c_{(3)}$ as 
\begin{equation}
S_B = T\int\! d^3\xi\,\sqrt{-\det g_{B}}
        + T\int {}^*c_{(3)}~,
\label{fullsb}
\end{equation}
where, from Eqs.~(\ref{elfb}) and (\ref{dc3}), we may take 
\begin{eqnarray}
c_{(3)}=-\frac{1}{8z^3}\,dt \wedge dx_1  \wedge dx_2\,.
\end{eqnarray}

\medskip 

The quadratic part $S_F$ is given by 
\begin{eqnarray}
S_F&=&
\frac{T}{2} \int\! d^3\xi\, \sqrt{-\det g_{B}} \,
g_B^{ij}g_{Fij}~
+\frac{T}{2} \int\! d^3\xi \, 
\epsilon^{ijk}
e^A_ie^B_j \bar\theta \Gamma_{AB} D_k\theta~.
\end{eqnarray}
This seems to have a bit unusual form.  Although one may
proceed with it, it is useful to rewrite the action so 
that the structure is manifest and understandable as much as
possible.  In order to rewrite the action, let us define the 
following quantities: 
\begin{eqnarray}
\Gamma_i \equiv e^A_i \Gamma_A\,, \qquad
\tilde\Gamma\equiv  \frac{1}{3!}\epsilon^{ijk}\Gamma_{ijk}\,.
\end{eqnarray}
Then it is an easy task to check that
\begin{eqnarray}
\frac{1}{2}\epsilon^{ijk}\Gamma_{ij}
=\Gamma^k \tilde\Gamma\,, \qquad
\Gamma^i=g^{ij}_B\Gamma_j \,.
\end{eqnarray}
By exploiting these formulae related to $\Gamma_i$ and introducing 
another quantity
\begin{equation}
\Gamma \equiv \frac{\tilde\Gamma}{\sqrt{-\det g_{B}}} \,,
\end{equation}
the action $S_{F}$ can be rewritten as 
\begin{eqnarray}
S_F &=&
-T\int\! d^3\xi~\sqrt{-\det g_{B}}\,
\bar\theta\Gamma^i(1-\Gamma)D_i\theta\,. 
\label{fullsf}
\end{eqnarray}
This is a form with the clearer structure as desired.  

\medskip 

Before closing this section, as one important remark, let us notice the presence 
of $1 - \Gamma$ in the action. 
As particular properties of $\Gamma$, one can show that $\Gamma^2=1$ 
with $\tilde\Gamma^2=-\det g_{B}$ and $\tr \Gamma=0$\,.  
This implies that half of the 32 components of $\theta$ are redundant
in the action and decoupled from the other dynamical variables.  
Indeed, this is nothing but the consequence of the
$\kappa$ symmetry that the M2-brane action possesses.

\section{Semiclassical analysis of an AdS$_2\times$S$^1$-brane} 

We consider a classical configuration of a single M2-brane
embedded in AdS$_4\times$S$^7/\mathbb{Z}_k$\,. It may be considered as the dual to a Wilson line of the boundary superconformal field theory. We study the quadratic action describing the fluctuations around it. 
 
\subsection{Static Classical Solution}

Let us consider the static configuration,
\begin{gather}
t=\tau \,, \quad z=\sigma \,, \quad y=\rho \,, \quad
x_1=x_2=w^m=0\,, 
\label{clsol}\\
\theta=0 \,.
\label{clsoltheta}
\end{gather}
This configuration satisfies the 
equation of motion for a single M2-brane and preserves half the 
supersymmetries of the background geometry. Note that the equation 
of motion itself is the one derived from the purely bosonic action
(\ref{fullsb}) because $\theta =0$ for the above configuration. 
The world-volume of the M2-brane described by the configuration touches 
the AdS$_4$ boundary on which the three-dimensional superconformal
field theory lives. 

\medskip 

When remembering the case of AdS$_5$/CFT$_4$, the boundary of 
this kind of static configuration represents a Wilson line. Similarly, it would 
be natural to regard the boundary of our configuration as a Wilson line 
in the boundary theory. In fact, the configuration described by (\ref{clsol}) 
and (\ref{clsoltheta}) is reduced to a 1/2 BPS string world-sheet 
in type IIA string theory through the dimensional reduction. Then it is argued that 
the resulting configuration would correspond to a Wilson loop 
in the context of AdS$_4$/CFT$_3$ based on the ABJM model. 
In other words, the configuration given by (\ref{clsol}) 
and (\ref{clsoltheta}) can be regarded as the up-lift of 
the type IIA string world-sheet to M-theory. According to this up-lift,   
the Wilson loop gets corrections by KK modes and hence there might be a
possibility that it cannot be understood as a Wilson loop any more.
However, it is supersymmetric and so it possibly remains understandable 
as the Wilson loop even in the M-theory limit.

\medskip

Next we shall consider fluctuations around the configuration given by (\ref{clsol}) 
and (\ref{clsoltheta}). We will deal with the bosonic and fermionic fluctuations separately
in the following two subsections.

\subsection{Bosonic fluctuations}

Assuming that the bosonic fluctuations are transverse to the static 
configuration (\ref{clsol}), we consider the following expansion of the fields: 
\begin{eqnarray}
t=\tau~,~~~
x_1=0+\tilde x_1~,~~~
x_2=0+\tilde x_2~,~~~
z=\sigma~,~~~
w^m = 0 + \zeta^m~,~~~
y=\rho \,.
\label{bosonic fluc}
\end{eqnarray}
Here the fields, $\tilde x_1$, $\tilde x_2$, $\zeta$, and $\bar\zeta$,
denote the bosonic fluctuations.  Because the classical configuration
(\ref{clsol}) plays the role in 
fixing the world-volume
diffeomorphism, there are no fluctuations along the world-volume.

\medskip 

Then we obtain that 
\begin{eqnarray}
S_B = S_B^{(0)} + S_B^{(1)} + S_B^{(2)} + \cdots\,, 
\end{eqnarray}
where $S^{(n)}_B$ is the part of the action containing the terms of
the $n$-th order of the bosonic fluctuations.  
Note that $S_B^{(1)}$ vanishes due to the equations of motion.  
The zeroth-order part is just the action for the classical 
configuration and is computed as
\[
S_B^{(0)} = T\int\! d^3\xi\, \frac{1}{4k \sigma^2} \,.
\]
Although the value of this action leads to the divergent contribution proportional to the volume,
it can be eliminated by the Legendre transformation as discussed in \cite{DGO} 
even for the M2-brane case. 

Before proceeding further, let us see the absence of such divergence by 
following the prescription of \cite{DGO}. We begin with a 
variation of the bosonic M2-brane action, $S_B$ of 
(\ref{fullsb}), which is obtained as
\begin{eqnarray}
\delta S_B &=&T\int\!\! d^3 \xi\,\partial_\sigma\left( \delta  X^M P_M^\sigma \right)
=\left.T\int\!\! d\tau d\rho \left( \delta  X^M P_M^\sigma \right)\right|_{\sigma=0}~.
\end{eqnarray}
The total derivative terms with respect to $\tau$ and $\rho$ vanish 
while the one with respect to $\sigma$ remains since the string world-sheet has the boundary 
at $\sigma =0$\,.   
In this variation we have used the equation of motion
\begin{eqnarray}
\frac{\partial \CL}{\partial X^M} - \partial_i\left(\frac{\partial \CL}{\partial(\partial_i X^M)}\right)=0
\end{eqnarray}
and defined $P_M^\sigma$ by
\begin{eqnarray}
P_M^\sigma \equiv \frac{\partial \CL}{\partial(\partial_\sigma X^M)}~.
\end{eqnarray}
The result implies that the action $S_{B}$ is a function of $X^M$ on the boundary.
By the way, the Wilson line is a function of $X^\mu=(t,x_1,x_2)$,
not of $Y^{\mu'}=(z,w^m,y)$.  
In order to make a connection with 
the Wilson line, we consider the Legendre transformation
$S'=S_B+S_L$ with
\begin{eqnarray}
S_L &=& - T\int\!\! d\tau d\rho~Y^{\mu'} P_{\mu'}^\sigma \,.
\end{eqnarray}
The variation of the transformed action is obtained as
\begin{eqnarray}
\delta S'&=&T\int\!\! d\tau d\rho~(\delta X^\mu P_\mu^\sigma 
+ Y^{\mu'}\delta P_{\mu'}^\sigma)\Big|_{\sigma=0}\,. 
\end{eqnarray}
This means that the action $S'$ is a function of $X^\mu$ and 
$P_{\mu'}^\sigma$ and thus we see that $S'$ is more appropriate 
to examine the correspondence to the Wilson line.  

\medskip

Having the suitable action $S'$, we are ready to investigate  
the divergence structure, which is basically given by evaluating 
$S_B$ and $S_L$.  The value of $S_L$ is evaluated in our static gauge as
\begin{eqnarray}
S_L=- T\int\!\! d\tau d\rho~ z \frac{\partial_\sigma z}{4k z^2}\Big|_{\sigma=\epsilon}
= -T\int \!\!d\tau d\rho\, \frac{1}{4k\epsilon}~,
\end{eqnarray}
where taking the limit of $\epsilon \rightarrow 0$ is assumed
implicitly. On the other hand, the divergent contribution from $S_B^{(0)}$ is 
evaluated as
\begin{eqnarray}
S_B^{(0)}=T\int\!\! d^3\xi\, \frac{1}{4k \sigma^2} 
=T\int\!\! d\tau d\rho \left(-\frac{1}{4k \sigma}\right)
  \Big|_{\sigma=\epsilon}^{\sigma=\infty}
=+T\int\!\! d\tau d\rho~\frac{1}{4k\epsilon }~.
\end{eqnarray}
Thus the divergence from $S_B^{(0)}$ and $S_L$ cancels out 
each other, and the volume-divergence in $S_B^{(0)}$ can be eliminated.
 

\medskip 

Now let us return to our main concern, that is, $S_B^{(2)}$.
In order to extract $S_B^{(2)}$, we first note that
the K\"ahler form $F$ is expanded as
\begin{eqnarray}
F=\frac{i}{2}d\zeta\wedge d\bar{\zeta}
+\cdots~,
\label{F expansion}
\end{eqnarray}
and the one-form potential $A$ becomes 
\begin{eqnarray}
A=\frac{i}{2}(\zeta d\bar{\zeta}-\bar{\zeta} d {\zeta})
+\cdots\,, \nonumber 
\end{eqnarray}
where ``$\cdots$'' represents the terms with higher-order in 
fluctuations (higher than the quadratic order). 
Then the metric $g_{Bij}$ is expanded as follows:
\begin{eqnarray}
g_{Bij}&=&g_{Bij}^{(0)}
+g_{Bij}^{(2)}+\cdots~,
\nonumber\\
g_{Bij}^{(0)}&=&
\diag\left(-\frac{1}{4\sigma^2},\frac{1}{4\sigma^2},\frac{1}{k^2}\right)~,
\cr
g_{Bij}^{(2)}&=&\frac{1}{4\sigma^2}\partial_{(i} \chi \partial_{j)}\bar\chi
+\partial_{(i}\zeta^m\partial_{j)}\bar\zeta^m
+\frac{i}{k}\partial_{(i}\rho (\zeta\partial_{j)}\bar\zeta-\bar\zeta\partial_{j)}\zeta)
~,
\label{g_B(0)}
\end{eqnarray}
where $\tilde x_1$ and $\tilde x_2$ have been combined to form a
complex field $\chi$\,,
\begin{eqnarray}
\chi=\tilde x_1+i \tilde x_2\,.  \nonumber 
\end{eqnarray}
The expanded metric subsequently allows us to expand the
determinant of the metric as
\begin{eqnarray}
\sqrt{-\det g_B}&=&\sqrt{-\det g_B^{(0)}}\sqrt{\det(1+(g_B^{(0)})^{-1}g_B^{(2)}+\cdots)}
\cr
&=&\sqrt{-\det g_B^{(0)}}\left(1+\frac{1}{2}
\tr ((g_B^{(0)})^{-1}g_B^{(2)})\right)+\cdots\,. \nonumber 
\end{eqnarray}

\medskip 

We are now ready to write down the quadratic action $S_B^{(2)}$.
For the NG action, we find that the quadratic part is given by 
\begin{align}
(S_\mathrm{NG})_B^{(2)} &= TR^3\int\!\! d^3\xi\,
\sqrt{-\det g_B^{(0)}}\left[
\frac{1}{2}g_B^{(0)ij}
\left(\frac{1}{4\sigma^2}\partial_i \chi \partial_j\bar\chi
+\partial_i\zeta^m\partial_j\bar\zeta^m
\right)
+\frac{i}{2}k(\zeta\partial_\rho\bar\zeta-\bar\zeta\partial_\rho\zeta)
\right]\,. \nonumber 
\end{align}
The contribution from the WZ term becomes 
\begin{eqnarray}
(S_\mathrm{WZ})_B^{(2)}
&=& - TR^3 \int\! d^3\xi\, 
\frac{3i}{32\sigma^4}(\chi\partial_\rho \bar\chi
-\bar\chi\partial_\rho \chi)  \,,
\label{wzb2}
\end{eqnarray}
with the help of integration by parts. 

\medskip

At this point, there is a comment on
the overall factor $TR^3$\,.  
It can be absorbed into the fluctuations by the redefinition, 
\begin{equation}
\sqrt{TR^3}\,\chi\to\chi\,, \qquad \sqrt{TR^3}\,\zeta\to\zeta\,, 
\label{redef}
\end{equation}
and disappears from the action.
As for the higher-order contributions $S^{(n>2)}$\,, however,  
we have inverse powers of the factor after the redefinition.
Thus, by take the limit  $TR^3\to\infty$, all such 
contributions simply vanish and what remains is the contributions
up to the quadratic order.

\medskip 

After the redefinition (\ref{redef}), 
the quadratic action for the bosonic part 
is obtained as
\begin{eqnarray}
S_B^{(2)} &=& (S_\mathrm{NG})_B^{(2)} + (S_\mathrm{WZ})_B^{(2)}  \nonumber \\ 
&=& \int\!\! d^3\xi\,
\sqrt{-\det g_B^{(0)}}
\Bigg[
\frac{1}{2}g_B^{(0)ij}
\left(\frac{1}{4\sigma^2}\partial_i \chi \partial_j\bar\chi
+\partial_i\zeta^m\partial_j\bar\zeta^m
\right) 
\nonumber \\
&& \hspace*{3.5cm}
+\frac{i}{2}k(\zeta\partial_\rho\bar\zeta-\bar\zeta\partial_\rho\zeta)
-
\frac{3i}{8}k\frac{1}{\sigma^2}
(\chi\partial_\rho \bar\chi
-
\bar\chi\partial_\rho\chi)
\Bigg]\,.
\end{eqnarray}
By introducing a new field $\eta$ defined as 
\begin{eqnarray}
\eta \equiv \frac{1}{2\sigma}\chi\,,
\label{redefinition}
\end{eqnarray}
we obtain the canonical action given by  
\begin{eqnarray}
S_B^{(2)}=\frac{1}{2}\int\! d^3\xi\,
\sqrt{-\det g_B^{(0)}}& \Bigg[&
g_B^{(0)ij}
\left(\partial_i \eta \partial_j\bar\eta
+\partial_i\zeta^m\partial_j\bar\zeta^m
\right)
+8\eta\bar\eta
\cr&&
+{i}k(\zeta\partial_\rho\bar\zeta-\bar\zeta\partial_\rho\zeta)
- 3ik (\eta\partial_\rho \bar\eta
- \bar\eta\partial_\rho \eta)
\Bigg] \,.
\label{bosonic fluc2}
\end{eqnarray}
The relation (\ref{redefinition}) will be important also when we discuss its behavior 
near the boundary.

\subsection{Fermionic fluctuations}

Next we consider the fermionic fluctuations around the solution (\ref{clsol}) and 
(\ref{clsoltheta}). Since the classical value of $\theta$ is zero,  
the action for the fermionic fluctuations is obtained simply by substituting 
(\ref{clsol}) into the action in $S_F$ (\ref{fullsf}) and regarding $\theta$ as the fluctuation.  

\medskip 

For notational clarity, we first put a bar to the ingredients of 
$S_F$ evaluated at the classical solution (\ref{clsol}), namely
\begin{eqnarray}
&&
\bar\Gamma_i=(\frac{1}{2\sigma}\Gamma_0,\frac{1}{2\sigma}\Gamma_3,\frac{1}{k}\Gamma_\natural)~,~~~
\bar\Gamma^i=(-{2\sigma}\Gamma_0,{2\sigma}\Gamma_3,{k}\Gamma_\natural)~,~~~
\cr
&&
\bar w^{03}_\tau=-\frac{1}{\sigma}~,~~
\bar w^{13}=\bar w^{23}=0~,~~
\bar w_\rho^{\tilde a\tilde b}=-F^{\tilde a\tilde b}\frac{1}{k}~,~~
\bar w^{\natural \tilde a}=0~.
\end{eqnarray}
Then the covariant derivative becomes 
\begin{eqnarray}
\bar D_\tau\theta&=&\partial_\tau\theta -\frac{1}{2\sigma}\Gamma_{03}\theta
+\frac{1}{2\sigma}\Gamma^{123}\theta~,~~~
\nonumber\\
\bar D_\sigma\theta&=&
\partial_\sigma\theta-\frac{1}{2\sigma}\Gamma^{012}\theta~,~~~
\\
\bar D_\rho\theta&=&
\partial_\rho\theta-\frac{1}{2k}\Gamma^{0123}{}_\natural \theta
-\frac{1}{4k}F^{\tilde a\tilde b} \Gamma_{\tilde a\tilde b}\theta~.
\nonumber
\end{eqnarray}
Because the action $S_F$ in (\ref{fullsf}) is already quadratic in $\theta$,
the action evaluated at the classical solution (\ref{clsol}) is
the desired one, that is, $S_F^{(2)}$.  As briefly mentioned just after (\ref{fullsf}), 
the M2-brane action has the fermionic$\kappa$ symmetry, which should be fixed before doing any practical
calculation with the action.  Here, from the $\kappa$ symmetry
transformation rule for $\theta$, 
\[
\delta_\kappa \theta = (1+\Gamma){\kappa} 
\]
and the form of $S_F$, we simply take the $\kappa$ symmetry fixing
condition as
\begin{equation}
(1 + \Gamma ) \theta = 0 \,, \label{fixing}
\end{equation}
which is used usually when one considers the number of supersymmetries
preserved by a brane configuration in a given supersymmetric background. Then,
with the fixing condition (\ref{fixing}), the quadratic action for the fermionic
part is obtained as
\begin{eqnarray}
S_F^{(2)}=
-2\int\!\! d^3\xi~\sqrt{-\det g_B^{(0)}}
&\Big[&
\bar\theta_{-}
\bar\Gamma^\tau
(\partial_\tau-\frac{1}{2\sigma}\Gamma_{03})\theta_{-}
+\bar\theta_{-}
\bar\Gamma^\sigma
\partial_\sigma\theta_{-}
+\bar\theta_{-}
\bar\Gamma^\rho
\partial_\rho\theta_{-}
\cr&&
+\frac{3}{2}\bar\theta_{-}\Gamma^{0123}\theta_{-}
-\frac{1}{2}\bar\theta_{-}
 \Gamma_\natural(\Gamma_{45}+\Gamma_{67}+\Gamma_{89})\theta_{-}
\Big]~, \label{fermion-11D}
\end{eqnarray}
where we have introduced $\theta_-$ defined as 
\begin{eqnarray}
\theta_{-} \equiv P_{-}\theta\,, \qquad 
P_{-}=\frac{1}{2}(1 - \Gamma_{03\natural})\,,
\end{eqnarray}
and the overall factor $TR^3$ has been absorbed into 
$\theta_{-}$ by the redefinition 
$\sqrt{TR^3}\,\theta_{-}\to\theta_{-}$\,.  
After the redefinition as in the bosonic case, 
the higher-order contributions in fluctuations vanish in the limit $TR^3\to\infty$\,. 

\medskip

Since we are studying the fluctuations on the three-dimensional
world-volume of M2-brane, it is convenient to rewrite this 
action in such a way that makes the three-dimensional
structure manifest.  What we should do first is to take a suitable 
representation of the gamma matrices with the manifest $SO(1,2) \times SO(8)$
structure. A possible representation 
is the following, 
\begin{align}
&\Gamma_0= \rho_0\otimes\gamma_9~,
&&
\gamma_9=\gamma_{1\cdots 8}=\sigma_3\otimes \sigma_3\otimes \sigma_3 \otimes \sigma_3~,\cr
&\Gamma_3= \rho_1\otimes \gamma_9~,
&&\rho_\alpha=(i\sigma_2,\sigma_1,\sigma_3)~,
\cr
&\Gamma_\natural= \rho_2 \otimes \gamma_9~,
&&
\cr
&\Gamma_1=1\otimes\gamma_1~,
&&
\gamma_1=\sigma_1\otimes \sigma_3\otimes \sigma_3 \otimes \sigma_3~,~~~
\cr
&\Gamma_2=1\otimes\gamma_2~,
&&
\gamma_2=\sigma_2\otimes \sigma_3\otimes \sigma_3 \otimes \sigma_3~,~~~
\cr
&
\Gamma_4=1\otimes\gamma_3~,
&&
\gamma_3=1\otimes \sigma_1\otimes \sigma_3 \otimes \sigma_3~,~~~
\cr
&
\Gamma_5=1\otimes\gamma_4~,
&&
\gamma_4=1\otimes \sigma_2\otimes \sigma_3 \otimes \sigma_3~,~~~
\cr
&\Gamma_6=1\otimes\gamma_5~,
&&
\gamma_5=1\otimes 1\otimes \sigma_1 \otimes \sigma_3~,~~~
\cr
&
\Gamma_7=1\otimes\gamma_6~,
&&
\gamma_6=1\otimes 1\otimes \sigma_2 \otimes \sigma_3~,~~~
\cr
&
\Gamma_8=1\otimes\gamma_7~,
&&
\gamma_7=1\otimes 1\otimes 1 \otimes \sigma_1~,~~~
\cr
&
\Gamma_9=1\otimes\gamma_8~,
&&
\gamma_8=1\otimes 1\otimes 1 \otimes \sigma_2~,~~~
\label{Gamma matrices}
\end{align}
where $\rho_\alpha$ are the gamma matrices in three dimensions, and
$\{\gamma_1,\cdots,\gamma_8\}$ are the $SO(8)$ gamma matrices.  
In this representation, the $B$-conjugation matrix $B$
is defined as 
\[
\Gamma_A^*=+B\Gamma_A B^{-1}\,,
\]
and satisfy the relations $B^\dag B=1$ and $B^T=B$\,. 
The matrix $B$ may be represented by 
\begin{eqnarray}
B=\Gamma_{2579}=1\otimes i\sigma_2\otimes \sigma_1\otimes i\sigma_2\otimes\sigma_1\,.
\label{B matrix}
\end{eqnarray}
Then the charge conjugation matrix $C=BA^\dag$ with $A=\Gamma_0$
is given by
\begin{eqnarray}
C=-\Gamma_{02579}=\rho_0\otimes \sigma_1\otimes\sigma_2\otimes\sigma_1\otimes\sigma_2\,.
\label{C matrix}
\end{eqnarray}
By using this representation,
the action reduces to
\begin{eqnarray}
S_F^{(2)}&=&-2\int\!\!d^3\xi\, \sqrt{-\det g_B^{(0)}}\bigg[
\theta_{-}^T \rho_0 \rho^i\otimes \sigma_1\otimes\sigma_2\otimes\sigma_1\otimes\sigma_2\,\nabla_i \theta_{-}\cr&&
\qquad \qquad \quad -\frac{i}{2}\theta_{-}^T\rho_0\rho_2\otimes
\Big(
3\sigma_1\sigma_3\otimes \sigma_2\otimes \sigma_1\otimes \sigma_2
+ \sigma_1\otimes \sigma_2\sigma_3\otimes \sigma_1\otimes \sigma_2
\cr&& \qquad \qquad \quad  
+ \sigma_1\otimes \sigma_2\otimes \sigma_1\sigma_3\otimes \sigma_2
+ \sigma_1\otimes \sigma_2\otimes \sigma_1\otimes \sigma_2\sigma_3
\Big)\theta_{-}
\bigg]~.
\end{eqnarray}
Here we have introduced the spinor covariant derivative $\nabla_i$ on
the world-volume as follows.  
From $g^{(0)}_{Bij}$ in (\ref{g_B(0)}), the dreibein and the spin connection 
are obtained as
\begin{eqnarray}
\hat e^\alpha=\left(
\frac{d\tau}{2\sigma},
\frac{d\sigma}{2\sigma},
\frac{d\rho}{k}
\right) \,, \quad
\hat w^0{}_1=-\frac{d\tau}{\sigma} \,.
\end{eqnarray}
Then the spinor covariant derivative
on the world-volume \adss{2}{1} is given by
\begin{eqnarray}
&&
\nabla_i=\partial_i+\frac{1}{4}\hat w_i^{\alpha\beta}\rho_{\alpha\beta}
=\left(\partial_\tau-\frac{1}{2\sigma}\rho_2,\partial_\sigma,\partial_\rho
\right) \,,
\cr&&
\rho_i=\hat e^\alpha \rho_\alpha \,, \quad
\rho^i= g_B^{(0)ij}\rho_j=\left(
-2\sigma\rho_0,
2\sigma\rho_1,
k\rho_2
\right) \,.
\end{eqnarray}

\medskip 

In order to describe the spinors from the M2-brane world-volume
perspective,
we decompose
the 32-component spinor $\theta$ into 16 two-component complex 
spinors $\vartheta^{\alpha_1\cdots\alpha_4}$ ($\alpha_i=\pm$)
where the four index $(\alpha_1,\alpha_2,\alpha_3,\alpha_4)$ represent 
the four U(1) charges of U(1)$^4\subset $ SO(8).
We will see soon that the Majorana condition relates a half of complex spinors
to another half of them by complex conjugation.
Since the projector is represented by 
\begin{eqnarray}
&&
P_{-}=1\otimes \frac{1}{2}(1{-}\gamma_9)
=1\otimes \frac{1}{2}(1_{16}{-}\sigma_3\otimes\sigma_3\otimes\sigma_3\otimes\sigma_3
)~,
\end{eqnarray}
$\theta_-$ is composed of 8 
two-component complex spinors $\vartheta^{\alpha_1\cdots\alpha_4}$ with
$\alpha_1\alpha_2\alpha_3\alpha_4={-}1$, namely
\begin{eqnarray}
&&
\vartheta^{-+++}\,, \quad 
\vartheta^{+++-}\,, \quad 
\vartheta^{++-+}\,, \quad 
\vartheta^{+-++}\,, \quad 
\label{+ spinors} \\
&&
\vartheta^{+---}\,, \quad 
\vartheta^{---+}\,, \quad 
\vartheta^{--+-}\,, \quad 
\vartheta^{-+--}\,.
\label{- spinors}
\end{eqnarray}
In terms of the two-component spinors, the action is rewritten as 
\begin{eqnarray}
S_F^{(2)}&=&-2\int\!\! d^3\xi\, \sqrt{-\det g_B^{(0)}}\sum_{\alpha_1\alpha_2\alpha_3\alpha_4=-1}\bigg[
-\alpha_2\alpha_4(\vartheta^{-\alpha_1-\alpha_2-\alpha_3-\alpha_4})^T\rho_0 
\rho^i\nabla_i \vartheta^{\alpha_1\alpha_2\alpha_3\alpha_4}\cr&&
\qquad \qquad \quad 
+\frac{i}{2}\alpha_2\alpha_4(
3\alpha_1+\alpha_2+\alpha_3+\alpha_4
)(\vartheta^{-\alpha_1-\alpha_2-\alpha_3-\alpha_4})^T\rho_0 \rho_2
\vartheta^{\alpha_1\alpha_2\alpha_3\alpha_4}
\bigg]~,
\end{eqnarray}
where the summation is taken over the eight spinors in (\ref{+ spinors}) and (\ref{- spinors}).
To derive this expression, we have used $\sigma_3\vartheta^\alpha=\alpha\vartheta^\alpha$,
$\sigma_1\vartheta^\alpha=\vartheta^{-\alpha}$ and $\sigma_2\vartheta^\alpha=-i\alpha \vartheta^{-\alpha}$.

\medskip

One may think that there is a problem at this point.  
While $\theta_{-}$ has 16
independent real components, the naive counting of real
components for the eight spinors $\vartheta^{\alpha_1\cdots\alpha_4}$
gives 32.  Actually, this point is not problematic and cured by the 
Majorana condition, which relates the four spinors in (\ref{+ spinors}) 
to the four spinors in (\ref{- spinors}). The proof is as follows.
The charge conjugate of a spinor $\psi$ is given by
\begin{eqnarray}
\psi^c=B^{-1}\psi^*~ \,,
\end{eqnarray}
where $B$ is given in (\ref{B matrix}).
The Majorana condition is the equality $\psi^c=\psi$, and thus
the eleven dimensional Majorana spinor $\theta_+$ satisfies
\[
\theta_+\equiv B^{-1}\theta_+^* \,.
\] 
This implies immediately that 
\begin{eqnarray}
\vartheta^{\alpha_1\cdots\alpha_4}=\alpha_1\alpha_3(\vartheta^{-\alpha_1  -\alpha_2 -\alpha_3 -\alpha_4})^*\,. 
\label{reality} 
\end{eqnarray}
Thus the independent spinors are provided by 
(\ref{+ spinors}) or (\ref{- spinors}).  Hereafter, we will take the spinors in  
(\ref{+ spinors}) as the independent ones.

\medskip 

In fact, using the relations 
\[\alpha_2\alpha_4(\vartheta^{-\alpha_1-\alpha_2-\alpha_3-\alpha_4})^T\rho_0 
=\alpha_1\alpha_2\alpha_3\alpha_4(\vartheta^{\alpha_1\alpha_2\alpha_3\alpha_4})^\dag\rho_0 
= {-}(\vartheta^{\alpha_1\alpha_2\alpha_3\alpha_4})^\dag\rho_0\,,
\] 
we obtain the following action 
\begin{eqnarray}
S_F^{(2)} &=& 4\int\!\! d^3\xi\, \sqrt{-\det g_B^{(0)}}\sum_{(\ref{+ spinors})}\bigg[
\bar\vartheta^{\alpha_1\alpha_2\alpha_3\alpha_4}
\rho^i\nabla_i \vartheta^{\alpha_1\alpha_2\alpha_3\alpha_4}\cr&& 
\qquad \qquad \quad 
-\frac{i}{2}(
3\alpha_1+\alpha_2+\alpha_3+\alpha_4
)\bar\vartheta^{\alpha_1\alpha_2\alpha_3\alpha_4}\rho_2
\vartheta^{\alpha_1\alpha_2\alpha_3\alpha_4}
\bigg]~,
\label{action theta 3-dim}
\end{eqnarray}
where the bar denotes the Dirac conjugate: 
$\bar\vartheta\equiv \vartheta^\dag \rho_0^\dagger$.
This is obviously of the three-dimensional form and 
$\vartheta^{\alpha_1\cdots\alpha_4}$ are massive complex fermions
propagating on \adss{2}{1} with the metric $g^{(0)}_{B ij}$ of
(\ref{g_B(0)}).

\section{KK Reduction from AdS$_2\times$S$^1$ to AdS$_2$}

The M2-brane considered in the previous section is described
by the three-dimensional field theory on the AdS$_2\times$S$^1$
world-volume.  We now take $S^1$ as the M-theory circle, and 
perform the KK reduction.  Then the resulting two-dimensional theory
is defined on AdS$_2$ and describes the KK modes including the IIA
string excitations.  In this section, we investigate the KK
spectrum on the AdS$_2$ space, especially focusing on the
cases of $k=1,2$.

\subsection{KK Reduction of bosonic sector}


First note that the parameter $k$ can be absorbed into the definition of 
$\rho$ by $\rho/k\to\rho$\,, and then the period of $\rho$ becomes 
$2\pi/k$: $\rho\sim \rho +2\pi/k$.  For large $k$\,, the circle 
parametrized by $\rho$ shrinks and \adss{4}{7}$/\bbZ_k$ reduces to 
\adscp.  The conformal field theory on the boundary turns out to be the 
$\CN=6$ ABJM theory.

\medskip 

When $k$ is infinite, the only contribution is the zero modes along $\rho$ direction. 
Suppose that $\eta$ and $\zeta$ are independent from $\rho$\,,
say, $\eta=\eta_0(\tau,\sigma)$ and $\zeta=\zeta_0(\tau,\sigma)$\,. 
The bosonic quadratic action (\ref{bosonic fluc2}) is reduced to
\begin{eqnarray}
S_B^{(2)}=\frac{1}{2}\frac{2\pi}{k}\int\! d^2\xi
\sqrt{-\det g_0}&\Bigg[&
g_0^{\hat i\hat j}
\left(\partial_{\hat i} \eta_0 \partial_{\hat j}\bar\eta_0
+\partial_{\hat i}\zeta^m_0\partial_{\hat j}\bar\zeta^m_0
\right)
+2\eta_0\bar\eta_0
\Bigg]\,, \label{4.1}
\end{eqnarray}
where $\xi^{\hat i}=(\tau,\sigma)$ and $g_{0\hat i\hat j}$ is the
\ads{2} metric with the unit radius
\begin{eqnarray}
g_{0\hat i\hat j}=\diag (-\frac{1}{\sigma^2},\frac{1}{\sigma^2})\,.
\label{AdS2 unit radius}
\end{eqnarray}
The action (\ref{4.1}) contains one massive complex scalar $\eta_0$ 
with $m^2=2$ and three massless complex scalars $\zeta^m_0$ propagating 
on the \ads{2} world-sheet with the metric (\ref{AdS2 unit radius}).

\medskip

As for $k=1,2$\,, 
the zero-mode argument would be insufficient 
because the supersymmetry of the boundary 
conformal field theory enhances from $\CN=6$ to $\CN=8$\,. Since the
information along the circle is expected to be crucial in this 
supersymmetry enhancement, we should now take the non-zero KK modes into account.  

\medskip 

Let us expand the bosonic fields as
\begin{eqnarray}
\zeta=\sum_{p\in\bbZ} e^{ip\rho} \zeta_p(\tau,\sigma)\,, \qquad 
\eta=\sum_{q\in\bbZ} e^{iq\rho} \eta_q(\tau,\sigma)\,. \label{them}
\end{eqnarray}
Substituting (\ref{them}) into (\ref{bosonic fluc2}), we obtain
\begin{eqnarray}
S_B^{(2)}=
\frac{2\pi}{2k}\int\! d^2\xi\, \sqrt{-\det g_0}&\Bigg[&
\sum_p\left(
g_0^{\hat i\hat j}\partial_{\hat i}\zeta_p^m \partial_{\hat j}\bar \zeta_p^m
+\frac{1}{4}kp(kp+2)\zeta_p^m \bar\zeta_p^m
\right)
\cr&&
+
\sum_q\left(
g_0^{\hat i\hat j}
\partial_{\hat i}\eta_q \partial_{\hat j}\bar \eta_q
+\frac{1}{4}(k^2q^2 {-} 6 kq+8)\eta_q\bar\eta_q
\right)
\Bigg]\,.
\end{eqnarray}
In this action, $\zeta_p$ and $\eta_q$ are described as massive scalars propagating
on \ads{2}.

\medskip 

We first consider the KK spectrum of $\zeta_p$\,. 
The mass of $\zeta_p$ is
\begin{eqnarray}
m^2(\zeta_p)=\frac{1}{4}kp(kp+2)~.
\end{eqnarray}
The Breitenlohner and Freedman (BF) bound \cite{BF} for the AdS$_2$ case is given by 
\[4m^2\ge -1\,.
\] 
This inequality is satisfied for all $p$ and is saturated
for $kp=-1$\,.  Note that $m^2(\zeta_p)$ {is invariant} under 
\[
kp \to -kp-2\,,
\]  
which implies that $\zeta_p$ and $\zeta_{p'}$ have the same mass
if $kp$ and $kp'$ are related by $kp'=-kp-2$.  Interestingly, the pairing of this 
kind is possible only when
\begin{eqnarray}
p+p'=-\frac{2}{k}\in \bbZ\,.
\end{eqnarray}
That is, only $k=1$ and $2$ are possible.
For $k=1$\,, the mode $\zeta_{-1}$ is not paired and appears alone 
because $p'=-p-2$ for $p=p'=-1$\,.

\medskip 

As for the mode $\eta_q$\,, its mass is
\begin{eqnarray}
m^2(\eta_q)=\frac{1}{4}(kq{-}4)(kq{-}2)\,.
\end{eqnarray}
The BF bound is satisfied for all the values of $q$ and is saturated for $kq={+}3$\,.
The $m^2(\eta_q)$ is invariant under $kq \to -kq {+}6$\,.  Thus there
appear two modes with the same mass when $6/k \in \bbZ$\,, which
includes the cases with $k=1,2$\,.

\medskip 

The resulting bosonic spectrum for $k=1,2$ is summarized as follows.
\begin{eqnarray*}
  \begin{array}{|c||c|c|c|}
  \hline
m^2       & kp   & kq & \mbox{degeneracy} \\ \hline\hline
-1/4      & -1   & 3 & 8 \\ \hline
  0     & 0,-2   & 2,4 & 16 \\ \hline
  3/4    & 1,-3    & 1,5 & 16 \\ \hline
  2     & 2,-4    & 0,6 & 16 \\ \hline
  15/4     & 3,-5    & -1,7 & 16\\ \hline
  \vdots    & \vdots    &\vdots  &\vdots \\ \hline
  \end{array}
\end{eqnarray*}
We note that, for the case of $k=2$\,, only modes with $kp\in 2\bbZ$ and 
$kq\in 2\bbZ$ survive.  Because $\zeta^m_p$ give six real scalars and 
$\eta_q$ gives two real scalars, we have 8 real scalars for 
$m^2=-1/4$ and 16 real scalars for each of 
$m^2=0, 3/4,2, 15/4,\cdots$\,.

\subsection{KK Reduction of fermionic sector}

Let us turn to the KK spectrum for the fermions.


As in the bosonic case, the fermionic variables are expanded as
\begin{eqnarray}
\vartheta^{\alpha_1\cdots\alpha_4}=\sum_{r\in\bbZ}e^{i\rho r}
\vartheta^{\alpha_1\cdots\alpha_4}_r(\tau,\sigma)\,. 
\end{eqnarray}
By substituting this expansion into the action
(\ref{action theta 3-dim}), the KK reduced action
is 
\begin{eqnarray}
S_F^{(2)}=-8\pi\int \!d^2\xi\, \sqrt{-\det g_B^{(0)}}\sum_{(\ref{+ spinors})\atop r\in\bbZ}\bigg[
\bar\vartheta
_r
\rho^{\hat i}\nabla_{\hat i} \vartheta
_r
-\frac{i}{2}(
3\alpha_1+\alpha_2+\alpha_3+\alpha_4
-2kr
)\bar\vartheta
_r\rho_2
\vartheta
_r
\bigg]\,, \label{4.9}
\end{eqnarray}
where $\hat i=\tau,\sigma$.
We have suppressed the indices $\alpha_1,\cdots,\alpha_4$ for
notational simplicity. The action (\ref{4.9}) can be rewritten 
so that it describes the fermions propagating on AdS$_2$ with 
the metric (in unit radius) of (\ref{AdS2 unit radius}) 
\begin{eqnarray}
S_F^{(2)}&=& -\frac{4\pi}{k}
\int \!d^2\xi \sqrt{-\det g_0}\sum_{(\ref{+ spinors})\atop r\in\bbZ}\bigg[
\bar\vartheta
_r
\hat\rho^{\hat i}\nabla_{\hat i} \vartheta
_r
-i\mu_{r}
\bar\vartheta
_r\rho_2
\vartheta
_r
\bigg]\,,
\cr
\mu_r &\equiv&
\frac{1}{4}(3\alpha_1+\alpha_2+\alpha_3+\alpha_4-2kr)\,,
\label{2d fermion action}
\end{eqnarray}
where we have defined the following quantities
\begin{eqnarray}
\hat\rho_{\hat i} \equiv \left(
\frac{1}{\sigma}\rho_0,\frac{1}{\sigma}\rho_1
\right)\,, \qquad 
\hat\rho^{\hat i} \equiv \left(
-{\sigma}\rho_0,{\sigma}\rho_1
\right)\,.
\end{eqnarray}

\medskip

Here we note that the covariant derivative in (\ref{2d fermion action})
may be regarded as the usual derivatine: $\nabla_{\hat i}=(\partial_\tau,\partial_\sigma)$.
This is seen easilly from (\ref{fermion-11D}).
The term including $\bar\Gamma^\tau \Gamma_{03}$ vanishes
because $C\Gamma^0\Gamma_{03}=C\Gamma_{3}$ is symmetric.
And so we may examine (\ref{2d fermion action}) with 
the replacement $\nabla_{\hat i}=(\partial_\tau,\partial_\sigma)$
below.

We are interested in the fermion mass spectrum. 
To read off the fermion mass spectrum, it is convenient to 
decompose the fermionic variables into the components and scale them like 
\begin{eqnarray}
\vartheta^{\alpha_1\cdots\alpha_4}_r=\left(
  \begin{array}{c}
    \phi^{(+)\alpha_1\cdots\alpha_4}_r  \\
    \phi^{(-)\alpha_1\cdots\alpha_4}_r   \\
  \end{array}
\right) \,.
\label{thphi}
\end{eqnarray}
The equation of motion obtained from (\ref{2d fermion action}) is rewritten as 
the coupled system of the first order differential equations
\begin{eqnarray}
\sigma(-\partial_\tau+\partial_\sigma)\phi^{(-)}_r-i\mu_r \phi^{(+)}_r&=&0\,, \cr
\sigma(+\partial_\tau+\partial_\sigma)\phi^{(+)}_r+i\mu_r \phi^{(-)}_r&=&0\,.
\end{eqnarray}
Let us multiply 
$\sigma(\partial_\tau+\partial_\sigma)$ to the
first equation (and $\sigma(-\partial_\tau+\partial_\sigma)$ to
the second equation, similarly) and introduce $\phi^{(1,2)}_r$ defined by
\begin{eqnarray}
\phi^{(1)}_r \equiv \phi^{(-)}_r+i\phi^{(+)}_r\,, \qquad 
\phi^{(2)}_r \equiv \phi^{(-)}_r-i\phi^{(+)}_r\,. \label{rotation}
\end{eqnarray}
Note that $\phi^{(1)}_r$ is not the complex conjugate of $\phi^{(2)}_r$ 
because $\phi^{(+)}_r$ and $\phi^{(-)}_r$ are complex variables.
As a result, $\phi_r^{(1,2)}$
should satisfy
\begin{eqnarray}
g_0^{ij}\nabla_i\nabla_j \phi^{(1)}_r-\mu_r(\mu_r-1)\phi^{(1)}_r&=&0\,, \cr
g_0^{ij}\nabla_i\nabla_j \phi^{(2)}_r-\mu_r(\mu_r+1)\phi^{(2)}_r&=&0\,,
\end{eqnarray}
and thus we see that a one two-component complex spinor
$\vartheta_r^{\alpha_1\cdots\alpha_4}$ represents two massive 
two-dimensional fermions with masses $m_F^2=\mu_r(\mu_r-1)$ and 
$\mu_r(\mu_r+1)$, or equivalently
\begin{eqnarray}
m_F^2=\mu(\mu-1)\,, \qquad \mu=\pm\mu_r\,. \label{m-form}
\end{eqnarray}

\medskip 

By using this mass formula (\ref{m-form}), let us see the mass spectrum for each of the components 
$\vartheta_r^{\alpha_1\cdots\alpha_4}$\,. 
For $\vartheta_r^{-+++}$\,, 
since $\mu_r=-\frac{1}{2}kr$\,, the mass spectrum is given by 
\begin{eqnarray*}
  \begin{array}{|c|c|c|c|}\hline
 kr      & \mu_r   & m_F^2(\mu=\mu_r)    &m_F^2(\mu=-\mu_r)    \\\hline\hline
 -3      &  3/2  & 3/4   &15/4    \\
 -2     &   1 &  0  &   2 \\
 -1     &  1/2  & -1/4 & 3/4     \\
 0     &  0  &  0  &   0 \\
 1     &  -1/2  & 3/4  &  -1/4   \\
 2     &  -1  &   2 &   0 \\
 3     &   -3/2 &   15/4 & 3/4   \\\hline
  \end{array}
\end{eqnarray*}
while similarly, for $\vartheta^{+++-}_r$, $\vartheta^{++-+}_r$ and
$\vartheta^{+-++}_r~$, since $\mu_r=\frac{1}{2}(2-kr)$\,,
\begin{eqnarray*}
  \begin{array}{|c|c|c|c|}\hline
 kr    & \mu_r & m_F^2 (\mu=\mu_r) & m_F^2 (\mu=-\mu_r) \\ \hline\hline
 0     &  1  & 0   &2    \\
 1     &  1/2  & -1/4   &3/4    \\
 2     &   0 &  0  &   0 \\
 3     &  -1/2  & 3/4 & -1/4     \\
 4     &  -1  &  2  &   0 \\
 5     &  -3/2  & 15/4  &  3/4   \\
 6     &  -2  &   6 &   2 \\
\hline
  \end{array}
\end{eqnarray*}
Note that the modes with $kr\in 2\bbZ$ only survive for the $k=2$ case. 

\medskip 

Thus the spectrum of the fermionic fluctuations is summarized as follows: 
\begin{eqnarray*}
  \begin{array}{|c||c|c|}\hline
m_F^2      & \mu  & \mbox{degeneracy}   \\\hline\hline
 -1/4      &  1/2  & 8       \\\hline
  0      &  0 & 8       \\
     &  1  &  8   \\ \hline
 3/4     &  -1/2  & 8      \\
      &  3/2  &  8   \\ \hline 
 2     &  -1  & 8      \\
      &  2  &  8   \\ \hline
  \vdots    & \vdots    &\vdots   \\
\hline
  \end{array}
\end{eqnarray*}
Since $\theta_r^{\alpha_1\cdots\alpha_4}$ represents a pair of 
two-dimensional fermionic modes, there are 8 modes 
for $m^2=-\frac{1}{4}$ and 16 modes for each of
$m^2=0,\frac{3}{4},2,\frac{15}{4},\cdots$\,. The fermionic result nicely 
agrees with the bosonic one with respect to the value of mass squared 
and the degeneracy, as it should be.

\subsection{$\CN=1$ scalar supermultiplets on AdS$_2$}

An $\CN=1$ scalar supermultiplet on AdS$_2$ is composed of a boson
with mass $m_B$ and a fermion with ``mass'' $\hat m_F$
{(which is the coefficient of the fermion mass term
and is different from $m_F$ above)}, where $m_B$ and $\hat m_F$
are given by, respectively, (see for example \cite{Sakai:1984vm}) 
\begin{eqnarray}
m_B^2=\mu(\mu-1)\,, \qquad \hat m_F=\mu\,. 
\end{eqnarray}

\medskip 

Now we can see that the obtained fluctuations form the $\CN=1$ scalar
supermultiplets.  In fact, we can identify $\mu$ for each of the
scalars as follows:
\begin{eqnarray*}
  \begin{array}{|c||c|}
  \hline
 m_B^2  &   \mu^\#
     \\\hline\hline
-\frac{1}{4}& (1/2)^8   \\
0& 0^8,~1^8    \\
\frac{3}{4} &(3/2)^8,~ (-1/2)^8   \\
2 &2^8,~(-1)^8   \\
\frac{15}{4} &(5/2)^8,~(-3/2)^8   \\
6 &3^8,~(-2)^8    \\
 \vdots &\vdots   \\\hline
  \end{array}
\end{eqnarray*}
Here $\#$ denotes the degeneracy.  It follows that the fluctuations form the 
$\CN=1$ scalar supermultiplets: 8 supermultiplets for each
$\mu\in\bbZ/2$\,, for $k=1,2$.

\medskip 

As an important remark, they are invariant under SO(8).  
This may suggest $\CN=8$
supersymmetry of the fluctuations. In fact, it is known that five
supermultiplets with $\mu=1$ and three supermultiplets with $\mu=-1$ 
form an $\CN=8$ supermultiplet on \ads{2}. It is worth finding 
out a tower of $\CN=8$ supermultiplets in our results.

\section{Fluctuations near the boundary}

The next task is to discuss the boundary behavior of the fluctuations. 
Our purpose here is to identify the non-normalizable modes, which are relevant 
to the operator insertions into the Wilson loop operator in the field-theory side. 
The Wilson line we are concerned about is expected to be 
1/2 BPS in the $\CN=8$ Chern-Simons-matter theory, which is the ABJM 
theory with enhanced $\CN=8$ supersymmetry.

\medskip 

From the consistency with the equation of motion,  
we can read off the behavior of a massive scalar field $\Phi$ near the boundary. 
When assuming the behavior
\begin{eqnarray}
\Phi ~~\to~~ \sigma^\lambda\, \hat \Phi(\tau) \qquad \mbox{as}~~~
\sigma ~~\to~~ 0 \,,
\end{eqnarray}
the equation of motion is reduced to 
\begin{equation}
-\sigma^{\lambda+2}\,\partial_\tau^2 \hat \Phi 
+\sigma^\lambda\left[\lambda(\lambda+1) -m^2\right]\hat \Phi=0\,. \label{it}
\end{equation}
The consistency of (\ref{it}) requires the two conditions: 
\[
\begin{array}{cccl}
\mbox{(i)} & \qquad \lambda+2 ~>~ 0 & \qquad 
\mbox{or} & \qquad  \partial_\tau^2\hat \Phi= 0\,, \\ 
\mbox{(ii)} & \qquad \lambda ~> ~0 & \qquad 
\mbox{or} &  \qquad 
\lambda=\frac{1}{2}(1\pm\sqrt{1+4m^2})\equiv \lambda_\pm\,. 
\end{array}
\]

\medskip 

These conditions are satisfied for the modes with $\lambda>0$ but such
modes do not reach the boundary. Similarly, since
$\lambda_+>0$, 
the mode with $\lambda=\lambda_+$
does not reach the boundary.
So we consider the modes with $\lambda=\lambda_-$.
The relation between $\lambda_-$ and $m^2$ is listed for some values
below.
\begin{eqnarray*}
  \begin{array}{|c||c|c|c|c|c|c|c|}\hline
m^2       & -1/4 &0 &3/4 & 2 & 15/4 &\cdots   \\\hline
\lambda_- &1/2 &0 &-1/2 & -1 & -3/2 &\cdots   \\\hline
  \end{array}
\end{eqnarray*}

\medskip 

As for $\zeta_p$\,, the modes with $m^2=-1/4$ cannot reach the boundary
because
\[
\zeta_p ~~\to~~ \sigma^{1/2} \hat  \zeta_p(\tau)\,,
\]
which vanishes as $\sigma\to 0$\,. The modes with 
$m^2 \ge 3/4$ diverge as approaching the boundary and 
cannot be regarded as small fluctuations. Thus the semiclassical approximation 
is not reliable any more. The modes with $m^2=0$ correspond to the fluctuations 
that can reach the boundary consistently with the approximation. 

\medskip 

For the case of $\eta_q$\,, it is related to the fluctuation $\chi_q$
through the field redefinition (\ref{redefinition}).  
So the corresponding fluctuation $\chi_q$ behaves as
\[
\chi_q=2\sigma \eta_q ~~\to~~ \sigma^{\lambda_-+1}\hat\eta_q\,.
\]
The modes with $m^2\le 3/4$ cannot reach the boundary and those 
with $m^2\ge 15/4$ diverge.  The modes with $m^2=2$ represent the
fluctuations that can reach the boundary.

\medskip

From (\ref{thphi}) and (\ref{rotation}), 
the fermionic modes $\phi^{(1,2)}$ are related to $\vartheta$ as 
\[
\vartheta_r=
\left(
  \begin{array}{c}
    \frac{1}{2i}(\phi^{(1)}_r-\phi^{(2)}_r)   \\
    \frac{1}{2}(\phi^{(1)}_r+\phi^{(2)}_r)   \\
  \end{array}
\right)
\,.
\]
To see the behavior of $\vartheta$ near the boundary, 
let us examine that of 
$
\phi^{(1,2)}$.
Noting that $\lambda_-=\frac{1}{2}(1-|2\mu-1|)$ with
$\mu=\mu_r$ for $\phi^{(1)}_r$ and $\mu=-\mu_r$ for $\phi^{(2)}_r$, we obtain
\begin{eqnarray*}
\phi^{(1)}\to \sigma^{\frac{1}{2}(1-|2\mu_r-1|)}\hat\phi^{(1)}(\tau)~,~~~
\phi^{(2)}\to \sigma^{\frac{1}{2}(1-|2\mu_r+1|)}\hat\phi^{(2)}(\tau)~.
\end{eqnarray*}
It is straightforward to see that $\phi^{(1)}$ remains small (or zero) 
when $0\le \mu_r\le 1$, 
while $\phi^{(2)}$ does
when $-1\le\mu_r\le 0$.
The equallity is satisfied for the mode which remains small and non-zero at the boundary.
Namely, $\phi^{(1)}$ for $\mu_r=0,1$ and $\phi^{(2)}$ for $\mu_r=-1,0$
are the modes reachng the boundary.
These are massless $m_F^2=0$.
The fermion with $\mu_r=0$ behaves near the boundary as
\begin{eqnarray*}
\vartheta\to\left(
  \begin{array}{c}
    \frac{1}{2i}(\hat\phi^{(1)}-\hat\phi^{(2)})   \\
    \frac{1}{2}(\hat\phi^{(1)}+\hat\phi^{(1)})   \\
  \end{array}
\right)~.
\end{eqnarray*}
On the other hand, for the fermion with $\mu_r=1$, $\phi^{(1)}$ remains small but
$\phi^{(2)}$ goes beyond semiclassical approximation.
Similarly for a fermion with $\mu_r=-1$, $\phi^{(2)}$ remains small but
$\phi^{(1)}$ goes beyond semiclassical approximation.
As a result, the fermions reaching the boundary are the following:
the fermion with $\mu_r=0$, a half of the fermion with $\mu_r= 1$ 
and a half of the fermion with $\mu_r=-1$.

\medskip 

In summary, the following modes can reach the boundary:
\begin{itemize}
  \item $\zeta_p^m$ ($m=1,2,3$) with $kp=0,-2$
  \item $\eta_q$ with $kq=0, {+}6$
  \item $\vartheta_r^{-+++}$ with $kr=0$
  \item  $\phi^{(1)-+++}_r$ with $kr=-2$\\ $\phi^{(2)-+++}_r$ with $kr=2$
  \item  $\vartheta_r^{+++-}$, $\vartheta_r^{++-+}$ and $\vartheta_r^{+-++}$ with $kr=2$
  \item  $\phi_r^{(1)+++-}$, $\phi_r^{(1)++-+}$ and $\phi_r^{(1)+-++}$ with $kr=0$
\\
$\phi_r^{(2)+++-}$, $\phi_r^{(2)++-+}$ and $\phi_r^{(2)+-++}$ with $kr=4$

\end{itemize}

The consistency condition (i) is satisfied for all these modes. They
are 16 bosonic fluctuations and 16 fermionic ones, which can be
regarded as to deform the Wilson line in the boundary gauge theory. 
In particular, 8 bosons and 5 fermions are contained as zero modes.

\medskip  

Among the modes that can reach the boundary, the zero modes may be 
problematic
when we consider the case of $k>2$.  In this case, only the zero modes 
survive 
in type IIA limit, and the 5 fermionic zero modes seem weird. 
Probably, the reason would be related to the fixing condition for the kappa 
symmetry
and to the fact that 8 supersymmetries are inevitably broken in moving to 
type IIA
description.

Hence there is subtlety of how to pick up the fermionic degrees of freedom 
that should be removed by using the $\kappa$ symmetry. 
We have taken a possible fixing condition for the $\kappa$ symmetry, 
but it does not seem to be compatible with the one in type IIA 
Green-Schwarz string discussed in \cite{IIA1,IIA2}. 
A comment on this point is given in the next section.
There may be another appropriate gauge fixing procedure which enables 
the fermion zero modes to survive all the way to the boundary.
An optimistic possibility is that our result may be correct in some way. In order to explain 
the origin of the fermionic degrees of freedom contained in the 1/2 BPS Wilson loops, 
there may be a non-trivial mechanism concerning the fermion sector 
as the theory flows from M-theory region to type IIA string theory.

\medskip


As for the number of degrees of freedom in each of bosonic and
fermionic fluctuations,
our result 16 seems twice the expected number and one may
feel curious.  At present, we do not have clear understanding for the
redundant degrees of freedom. 
Those may agree with the results on 1/2 BPS Wilson loops recently discussed 
in \cite{Lee}, where the Wilson loops are discussed as the Higgsing in the ABJM model  
and then there appear the two multiplets correponding to ${\bf (N-1,1)}$ and ${\bf (1,N-1)}$ representations after the Higgsing. 
Since this property may persist in the $\CN=8$ case,
the multiplets may correspond to our results though we do not have any 
confirmation with the exact matching of the degrees of freedom. 
As another possibility, we may argue that those would probably correspond to the non-ABJM fields 
discussed in \cite{GR} because we have started from the eleven-dimensional 
$\mathcal{N}=8$ setup with $k=1,2$.

\section{Summary and Discussion}

In this paper we have considered semiclassical fluctuations around a
single M2-brane configuration on AdS$_4\times$S$^7/\mathbb{Z}_k$\,.
The configuration is static, 1/2 BPS and its shape is
AdS$_2\times$S$^1$\,.  We have investigated the Kaluza-Klein reduction
on S$^1$ and shown that the resulting fluctuations form an infinite 
set of $\mathcal{N}=1$ supermultiplets on AdS$_2$ {for $k=1,2$}\,. 
The SO(8) invariance of the spectrum may suggest
the $\mathcal{N}=8$ supersymmetry on AdS$_2$. 

\medskip 

We also have discussed the behavior of the fluctuations near the
boundary of AdS$_2$\,. As a result, it has been shown that 16 bosonic
fluctuations and 16 fermionic ones can reach the boundary without
spoiling the semiclassical approximation.  It seems twice the
expectation but we argue that the redundant degrees of freedom should
correspond to the non-ABJM fields because we have started from the
eleven-dimensional $\mathcal{N}=8$ setup with $k=1,2$\,.

\medskip 

We have noted a subtlety for the fixing condition of the kappa symmetry. 
In other words, this problem is 
translated to the choice of classical solution in a sense that 
our $\kappa$-symmetry fixing condition (\ref{fixing}) depends on $\Gamma$ 
whose 
form is fixed once we choose a classical configuration.
In the present case, $\Gamma$ in (\ref{fullsf}) is fixed by choosing the classical configuration (\ref{clsol}).
After the $\kappa$-gauge fixing we are left with 8 spinors in (\ref{+ spinors})
and (\ref{- spinors}).
Under the breaking SO(8)$\to$SU(4)$\times$U(1), the $8_s$ representation decomposes into
$1_2+1_{-2}+6_0$. Then $1_2$ and $1_{-2}$ correspond to
$\vartheta^{-+++}$ and $\vartheta^{+---}$, respectively, and  
$6_0$ consists of $\vartheta^{-+--}$, $\vartheta^{--+-}$, $\vartheta^{---+}$, $\vartheta^{+-++}$, $\vartheta^{++-+}$, $\vartheta^{+++-}$\footnote{
The U(1) charges $(s_1,s_2,s_3,s_4)$ in \cite{DPY}
correspond to $(\gamma_9\alpha_1,\alpha_2,\alpha_3,\alpha_4)$
with $\gamma_9=-1$ in our notation.
Our spinors are complex subject to the reality condition (\ref{reality})
and we may regard (\ref{+ spinors}) as indipendent ones.}. 
If we start from the partially $\kappa$-gauge fixed type IIA string action in \adscp \cite{IIA1,IIA2}, $\vartheta^{-+++}$ and $\vartheta^{+---}$ are absent from the outset.
On the other hand, our result contains these spinors (even for zero modes). 
Hence our result cannot be derived 
from the IIA perspective.
It is interesting to examine different M2-brane configurations from the one we considered here 
and to investigate the relation to type IIA theory.

\medskip

Our aim here was to confirm the correspondence between (dimensionally reduced) 
M2-brane world-volume and Wilson loop from the analysis of the fluctuations 
in the case of AdS$_4$/CFT$_3$ as in AdS$_5$/CFT$_4$. For this purpose 
the fermion zero-modes in our result look weird. 
This may be because the $\kappa$-gauge fixing condition forced by the choice of the classical solution is not compatible with the IIA string setup.
We hope that there may be a nice interpretation to support our result. 
In any case, at the present stage, it is hard to answer whether the 
correspondence should hold or not. We need much effort to give a definite
answer and it remains as a future problem.

\medskip 

Another way is to consider semiclassical fluctuations around the
AdS$_2$ solution by starting directly from type IIA string theory on
AdS$_4\times\mathbb{C}$P$^3$ rather than the M2-brane theory. 
It would be interesting to consider the
interpretation of the fluctuations as a small deformation of the 1/2 BPS
Wilson line \cite{supermatrix} or the 1/6 BPS Wilson line
\cite{DPY,Chen,RSY}.  We hope that we could report some results in this
direction in the near future \cite{future}.

\medskip 

There are some other open problems. One of them is to compute the
semiclassical partition function around the static M2-brane solution
as in the case of AdS$_5$/CFT$_4$ \cite{DGT}.  Another one is to
consider a Wilson loop in the $\mathcal{N}=8$ three-dimensional CFT like
the BLG theory \cite{BL,G}. This has not been clarified at all, and so
it would be nice if we could shed light on this issue from our
approach.

\section*{Acknowledgment}

The authors would like to thank Machiko Hatsuda and Ta-Sheng Tai 
for useful discussions. 
The work of MS and KY was supported in part by the Grant-in-Aid for
Scientific Research (19540324) from the Ministry of Education, Science
and Culture, Japan, and by the Grant-in-Aid for the Global COE Program
``The Next Generation of Physics, Spun from Universality and
Emergence'' from the Ministry of Education, Culture, Sports, Science
and Technology (MEXT) of Japan.  The work of HS was supported by the
National Research Foundation of Korea (NRF) grant funded by the Korea
government (MEST) through the Center for Quantum Spacetime (CQUeST) of
Sogang University with grant number 2005-0049409 and also supported by
the NRF grant funded by the Korean government with grant number
NRF-2008-331-C00071 and R01-2008-000-21026-0.


\end{document}